\documentclass{aa}
\usepackage{graphicx}
\usepackage{rotating}
\usepackage{amsmath}
\usepackage{txfonts}
\usepackage{natbib}
\bibpunct{(}{)}{;}{a}{}{,}
\begin{document}

\title{\textbf{GeV emission from short gamma-ray bursts: \\the case of GRB~081024B}}

\author{Alessandra Corsi\inst{1,2}\thanks{Current address: LIGO laboratory, California Institute of Technology, CA 91125 Pasadena, USA. E-mail: corsi@caltech.edu} \and Dafne Guetta\inst{3} \and Luigi Piro\inst{2}}
\institute{Universit\`a degli studi di Roma ``Sapienza'' and INFN-Sezione di Roma, Piazzale
Aldo Moro 2, 00185 Roma, Italy. \and INAF - Istituto di
Astrofisica Spaziale e Fisica Cosmica di Roma, Via Fosso del
Cavaliere 100, 00133 Roma, Italy. \and INAF - Osservatorio Astronomico di Roma,
Via Frascati 33, 00040 Monte Porzio Catone, Italy }
\offprints{alessandra.corsi@roma1.infn.it,
guetta@oa-roma.inaf.it, luigi.piro@iasf-roma.inaf.it}

\date{\today}

\abstract{}{ {We investigate whether the high energy tail detected by the Fermi/LAT for the
short \object{GRB 081024B} can be caused by synchrotron and self-Compton emission in the context of either the internal
or external shock models.}} {For the internal shock scenario, we explore the
possibility of generating the high energy photons directly by means of the synchrotron process, or 
inverse Compton emission in which target photons are synchrotron photons produced in internal shocks
taking place in a lately emitted shell (delayed internal shocks). 
In the external shock scenario, we test whether the high energy tail can be an
extension of the afterglow synchrotron emission, or alternatively the inverse Compton
component associated with the afterglow synchrotron photons.}
{For the internal shock scenario, we conclude that only an inverse Compton component from delayed internal shocks can explain the high energy tail that extends to the GeV range. In the external shock scenario, we show
that the high energy tail may be interpreted as synchrotron afterglow emission, if the slow cooling phase starts as early as a few seconds after the trigger. On the other hand, the observed high energy tail is consistent with an inverse Compton component of the afterglow in the fast cooling regime.
}{}

\keywords{Gamma rays: bursts -- X-rays: Individuals
(\object{GRB~081024B}) -- X-rays: bursts -- radiation mechanisms:
non-thermal}

\authorrunning{A. Corsi, D. Guetta, L. Piro}

\titlerunning{GeV emission from the short \object{GRB~081024B}}

\maketitle

\section{Introduction}
The detection by the AGILE and Fermi satellites of substantial high energy
emission from short gamma-ray bursts \cite[GRBs, ][]{Abdoart,Abdo2009,Giuliani2009},
has challenged our understanding of this type of bursts as a high energy
source. These results are surprising because our expectations before the launch of Fermi, were that high energy
emission was more likely detectable from long GRBs
\citep[see e.g.][]{science}, which have a higher equivalent
isotropic energy and interstellar medium (ISM) number density
\citep[][]{Nakar2007}. However, Fermi observations of \object{GRB
081024B} show a longer-lasting ($\sim 3$~s)
tail with a few photons in the GeV range following the main event
\citep[][]{GCN8407,Abdoart}. Motivated by this result, we
analyze the conditions under which Fermi observations can be
explained by the most popular theoretical models.

In the internal-external shock scenario of the fireball model
\cite[see e.g. ][]{Meszaros1993,Sari1998}, GRB prompt and
afterglow emissions are understood to be produced by particles
accelerated via shocks into an ultra-relativistic outflow (fireball)
released during the burst explosion. While the prompt emission is
related to shocks developing into the ejecta (internal shocks, IS),
the afterglow arises from the forward external shock (ES) propagating into the ISM.

Synchrotron emission by the accelerated electrons is typically
invoked as the main radiation mechanism. However, inverse Compton
emission (IC) may also play an important role. Some synchrotron
photons can Compton-scatter from the shock-accelerated
electrons, producing an additional IC component at higher
energies. This mechanism is also called synchrotron self-Compton
(SSC) as the electrons responsible for the synchrotron emission are
also responsible for the IC radiation. The ratio of IC-to-synchrotron luminosities is proportional to the
square root of the ratio of the electron ($\epsilon_e$) to
magnetic ($\epsilon_B$) energy densities behind the shock front.
When this ratio is significantly above unity, the electron cooling
rate via IC emission cannot be neglected.

The IC emission from IS has been considered in various
contexts \cite[e.g.
][]{Pa1996,Pilla1998,Ghisellini2000,PanaitescuMes2000,DaiLu2002,Guetta2003,Baring2004,Peer2004,Asano2007,Fan2008b,GalliGuetta2008,Li2008,Yu2009,Toma2010}.
Here we focus on the model presented by \citet{Guetta2003}
where high-energy emission from IS during the prompt
GRB is computed, for both the synchrotron and IC components, as a
function of two free parameters: the Lorentz factor $\Gamma$ and
the variability time $t_v$ of the central engine that emits the
outflow. We note however that the IS emission for GRBs has been the subject of extensive amount of literature \citep[e.g. ][]{Meszaros1994,Sari1997,Daigne1998,Pilla1998,Panaitescu1999,Beloborodov2000,Spada2000,Ramirez2000}, to which the reader is referred.
IC emission from the ES \cite[see e.g.][ and references
therein]{SariEsin2001,ZhangMes2001} has been invoked to explain
GRB X-ray afterglows displaying properties difficult to reconcile
with the simplest synchrotron-only afterglow scenario \cite[e.g.
][]{Wei1998,Wei2000,Harrison2001,Corsi2005,Corsi2006,Chandra2008},
or in the context of higher energy emission from GRBs, in view of
EGRET and Fermi/LAT capabilities and results \cite[see e.g.
][]{Asaf2005,Wang2006,Gou2007,Galli2007,You2007,Fan2008,Fan2008b,Galli2008,Wang2009,Fan09}.

The detection of GRB high-energy (MeV to GeV) emission by AGILE
and Fermi/LAT may be particularly relevant to probing the mechanisms
active during the prompt-to-afterglow transition phase, when IC
emission from both the IS and ES may be invoked, and observations
are needed to help discriminate between different models. In this
context, we consider the case of the short \object{GRB~081024B},
for which a high energy emission tail was detected by the
Fermi/LAT after the prompt phase. \citet{Zou2008} 
concluded that both the IS and ES scenarios may produce
emission peaking at GeV energies, in agreement with the
observations for this burst. In this paper, we extend the analysis of \citet{Zou2008}, by 
taking into account \object{GRB~081024B} data published by \citet{Abdoart}. 
In the IS scenario, we consider the
possibility that the $\sim$ GeV emission from \object{GRB 081024B}
is due to synchrotron or IC emission from a lately emitted shell. The observations are used to derive constraints on the IS
model parameters. 
For the ES scenario, we investigate whether the high energy tail is a 
simple extension to high energies of the afterglow synchrotron
emission, or the SSC component associated with the afterglow
synchrotron photons. The model is constrained by considering not only the IC peak energy, which was considered by
\citet{Zou2008}, but also its luminosity, thus providing a
more stringent estimate of its compatibility with the observations.
Both the late IS and ES scenarios can naturally account for a
 delay between the GRB trigger time and the longer-lasting
high energy tail. This is remarkable given
 that a delay has indeed been observed in some
other cases \citep[see e.g. ][]{science,Abdo2009}.

\section{Observations}
\label{osservazioni}
At 21:22:40.86 UT on 24 October 2008, the Fermi Gamma-ray Burst Monitor (GBM) triggered on \object{GRB 081024B}. The light curve of the burst was characterized by a narrow spike of about 0.1 s (hereafter interval a), followed by a longer pulse, of about 0.7 s (hereafter interval b; Abdo et al. 2010). There is no evidence of emission after 0.8 s in GBM detectors covering the 8 keV - 5 MeV energy range \citep{Abdoart}. An event with energy $3.1\pm0.2$ GeV was detected after 0.55 s, while a second event of $1.7\pm0.1$ GeV was detected after 2.18 s \citep{Abdoart}. A time-resolved spectral analysis was performed in intervals a, b, and one third interval (hereafter interval c) in-between 0.8 s and 2.9 s after the trigger. The best-fit spectra were obtained by simultaneously fitting the signal from the GBM detectors in the energy range 8 keV - 36 MeV, and the LAT detectors (selecting transient events above 100 MeV; Abdo et al.2010).  \\
In interval a, the best fit to the GBM data is obtained using a power law with a low energy spectral index of $\alpha=-1.03^{+0.23}_{-0.19}$ and exponential cutoff around $E_{peak}\sim 2.7$ MeV \citep[see the upper panel of Fig. 3 in][or the continuous line in our Fig. 1]{Abdoart}, though its value is only marginally constrained. The fluence in the 100 MeV - 10 GeV energy range was estimated to be  $<4\times10^{-10}$ erg/cm$^{2}$, while the fluence measured in the 20 keV - 2 MeV range was  $(1.7\pm0.3)\times10^{-7}$ erg/cm$^{2}$.\\
The emission during interval b was fit with a Band plus a power-law model, or an exponential cut-off power-law plus a power-law model. The first yielded best-fit parameter values of $\alpha= -1.03^{+0.17}_{-0.14}$, $\beta= -2.1^{+0.11}_{-0.14}$, and $E_{peak}=2.0^{+1.9}_{-1.0}$ MeV \citep[see the second panel from top of Fig. 3 in][or the dashed line in our Fig. 1]{Abdoart}. The second yielded best-fit values of $\alpha=-0.7^{+0.4}_{-0.3}$ and $E_{peak}=1.6^{+1.5}_{-0.6}$ MeV for the cutoff power-law component; and $\beta=-1.68^{+0.10}_{-0.06}$ for the power-law component \citep{Abdoart}.\\
Finally, during interval c, the emission is more accurately represented by a simple power-law, with a best-fit photon index of $\beta=-1.6^{+0.4}_{-0.1}$ \citep[see the lowest panel of Fig. 3 in][or the dot-dashed line in our Fig. 1]{Abdoart}. The fluence measured in the 20 keV to 2 MeV energy range during this interval was $(4.3\pm3.2)\times10^{-8}$ erg/cm$^{2}$, with most of the energy being emitted in the 100 MeV - 10 GeV range, for a measured fluence of $(4.0\pm2.4)\times10^{-7}$ erg/cm$^{2}$ \citep{Abdoart}.

\object{GRB081024B} also triggered the Suzaku Wide-band All-sky Monitor
(WAM, 50 keV - 5 MeV) at $T_0=21:22:40.526$ UT \cite[][]{GCN8444}.
The light curve showed a double-peaked structure with a $T_{90}$
duration of $\sim 0.4$ s.  The fluence in 100 - 1000 keV range was
$(2.7^{+0.7}_{-1.0})\times10^{-7}$~erg~cm$^{-2}$.  The peak flux
within $0.5$~s was $1.1^{+0.3}_{-0.5}$~photons~cm$^{-2}$~s$^{-1}$
in the same energy range. Preliminary results showed that at least
2 MeV photons were detected, and the time-averaged spectrum from
$T_0$ to $T_0+0.5$~s was well fitted by a single power law, with a
photon index of $-1.24^{+0.25}_{-0.19}$ \cite[][]{GCN8444}.

Swift XRT began observing the field of the Fermi-LAT  around
$70.3$ ks after the trigger \cite[][]{GCN8410}. Thanks to a series
of follow-up observations \cite[][]{GCN8416,GCN8454}, it was
possible to establish that none of the three sources could be
the GRB X-ray counterpart because they were not fading.

\begin{figure}
    \begin{center}
        \includegraphics[width=8.cm]{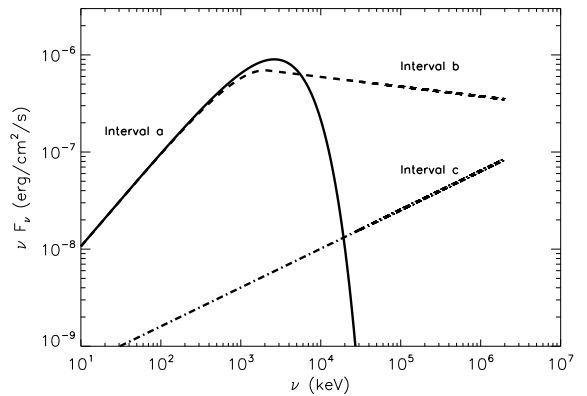}
        \caption{Best fit to the spectra of \object{GRB 081024B} during intervals a (continuous line - COMPT model), b (dashed line - Band model), and c (dot-dashed line - power-law model) as reported in Table 2 and Fig. 3 of \citet{Abdoart}.}
    \end{center}
\end{figure}

\section{The first 3 s of emission within the IS model}
\label{sec2}
The observed dichotomy in the spectral behavior of \object{GRB~081024B} during the first 3 s of emission, suggests that the properties of the central engine are evolving between interval a and c. During interval c, the observation of $\sim 2$ GeV photons implies an optically thin source in the GeV range, while during interval a the absence of emission above $\sim 10$ MeV and the unusually steep high energy photon index, suggest that the source is optically thick to pair production. Hereafter, we analyze in more detail this scenario, noting however that other explanations may also be invoked. For instance, an alternative possibility is that there is no emission at all in the GeV range: this would be the case if interval a is dominated by emission from a photosphere, rather than from an absorbed synchrotron spectrum. We refer the reader interested in this alternative explanation to papers such as e.g. Ioka (2010), Mizuta et al. (2010), Pe'er \& Ryde (2010), Toma et al. (2010), and references therein.

In the IS model \citep[e.g.][]{Guetta2003}, the central engine is supposed to emit a flow with Lorentz factor $\Gamma$, which is assumed to vary on a typical timescale $t_v$ (corresponding to an observed temporal variability of $\delta t_{obs}=(1+z)t_{v}$), with an amplitude $\delta \Gamma \sim \Gamma$. 
The shells collide at a radius $R \approx
2\Gamma^2ct_v=6 \times 10^{13} \Gamma^2_{2.5}t_{v,-2}$ cm, where
$\Gamma_{2.5}=\Gamma/10^{2.5}$ and $t_{v,-2}=t_v/(10^{-2}\rm~s)$. The
internal energy released in each collision is distributed among electrons, magnetic field, and
protons with fractions $\epsilon_e$, $\epsilon_B$, and $(1-\epsilon_e)$, respectively. The
electrons are accelerated in the shocks to a power-law distribution of energy $N(\gamma)
\propto \gamma^{-p}$, and radiatively cool by the combination of synchrotron and SSC
processes, the timescales of which are $t_{syn} \sim 6 \pi m_ec/\sigma_T B^{2}\gamma$ and
$t_{SSC}=t_{syn}/Y$, the combined cooling time being
$t_c=(1/t_{syn}+1/t_{SSC})^{-1}$=$t_{syn}/(1+Y)$, where $B$ is the magnetic field, and $Y$
is the Compton y-parameter \citep{sari96}$, Y \approx \epsilon_e/\epsilon_B$ for $\epsilon_e
<< \epsilon_B$ and $Y \approx (\epsilon_e/\epsilon_B)^{1/2}$ for $\epsilon_e
>> \epsilon_B$.

\subsection{Interval a: IS synchrotron emission from a compact source}

We now hypothesize that the lack of emission outside the GBM energy band (i.e. $E\gtrsim 30$ MeV) observed during interval a is due to the optical thickness for pair production. We assume that the \textit{unabsorbed} spectrum is a Band spectrum, of a low energy spectral slope of $\alpha\sim -1.03$ and peak energy $E_{peak}=2.7$ MeV as observed, but with a high-energy spectral slope of $\beta=-2.5$ \citep[as typically observed for GRB prompt spectra, see e.g.][]{Kaneko2006}. We note that a Band fit to the data during this interval poorly constrains $\beta$ to be less than $\sim -1.7$. The $\tau_{\gamma\gamma}$ for pair production is expressed as follows \citep[see e.g. ][]{Svensson1987,Lithwick2003}:

\begin{equation}
	\tau_{\gamma\gamma} (E)\sim \frac{0.1 \sigma_T N_{\gamma>E_{an}(E)}}{4\pi R^2},
\label{tau}
\end{equation}
where $\sigma_T$ is the Thompson cross-section, $R$ is the compactness of the source, and $N_{>E_{an}(E)}$ is the number of target photons, i.e. the number of photons with energy above $E_{an}$, where
\begin{equation}
	E_{an}(E)=\frac{(\Gamma m_e c^2)^2}{E (1+z)^2}=\frac{2.6\times10^{5}\Gamma^{2}}{(E/{\rm keV})(1+z)^2} {\rm keV}
	\label{ean}
\end{equation}
accounts for a photon of energy $E$ in the observer frame being attenuated by pair production by an interaction with softer photons, whose energy (also in the observer frame) is equal to or greater than $E_{an}(E)$. For a power-law spectrum of the form 
\begin{equation}
	N(E)=C (E/100 {\rm~ keV})^{\beta} \frac{\rm ph}{\rm cm^{2}s~keV},
\end{equation} 
one has
\begin{eqnarray}
	N_{\gamma>E_{an}(E)}=\frac{C ~4 \pi (d_L/{\rm cm})^{2}(\delta t_{obs}/{\rm s}) (E_{an}(E)/{\rm keV})^{1+\beta}}{-(1+\beta)(100)^{\beta}(1+z)^{2}}
	\label{np}
\end{eqnarray}
(where we are supposing $\beta<-1$). 

We define $E_{max}$ as the energy for which $\tau_{\gamma\gamma}(E_{max})=1$. Using $R=2 c \Gamma^2 \delta t_{obs}/(1+z)=6\times10^{10}\Gamma^2 \left[\delta t_{obs}/((1+z)\rm s)\right]$ cm, and substituting Eqs. (\ref{ean}) and (\ref{np}) into Eq. (\ref{tau}) we have
\begin{eqnarray}
\Gamma\sim\left[\frac{1.8\times10^{-47}~C (d_L/{\rm cm})^2  (2.6\times10^5)^{1+\beta}}{(1+z)^{(2+2\beta)}(100)^{\beta} (\delta t_{obs}/{\rm s})(-1-\beta)(E_{max}/{\rm keV})^{(1+\beta)} }\right]^{1/(2-2\beta)}.
\label{gammap}
\end{eqnarray}
No afterglow emission was detected for \object{GRB 081024B}, so the burst redshift is unknown. Hereafter we assume $z=0.1$ as a reference value for short GRBs, i.e. $d_{L}=1.4\times10^{27}$ cm for the luminosity distance. The Band spectrum is given by \citep{Band1993}:
\begin{eqnarray}
	N(E)=A \left(\frac{E}{100 {\rm~ keV}}\right)^{\alpha} e^{-E(2+\alpha)/E_{peak}}~~\frac{\rm ph}{\rm cm^{2}s~keV}\\\nonumber~~~~~{\rm for}~~~ E<\frac{(\alpha-\beta)E_{peak}}{(2+\alpha)},
\label{Bandlow}
\end{eqnarray} 
\begin{eqnarray}
	\nonumber N(E)=A \left(\frac{(\alpha-\beta)E_{peak}}{e(2+\alpha)100 {\rm~ keV}}\right)^{\alpha-\beta}\left(\frac{E}{100 {\rm~ keV}}\right)^{\beta} \frac{\rm ph}{\rm cm^{2}s~keV}=\\ =C_{Band}\left(\frac{E}{100 {\rm~ keV}}\right)^{\beta}~~\frac{\rm ph}{\rm cm^{2}s~keV} ~~~~~{\rm for} ~~~E>\frac{(\alpha-\beta)E_{peak}}{(2+\alpha)}\label{Bandhigh}.
\end{eqnarray} 

We note that these equations are obtained from Eq.~(1) of \citet{Band1993} by using $E_{peak}=(2+\alpha)E_0$  \citep[see e.g.][]{Piran1999}. We note also that the multiplicative factor $e^{\beta-\alpha}$ in Eq.~(1) of \citet{Band1993} is included in the first factor in parenthesis of the above equation. We can thus approximate the high energy portion of the \textit{unabsorbed} Band spectrum as \citep{Band1993}
\begin{eqnarray}
	N(E)=0.3~(E/100 {\rm~ keV})^{-2.5} \frac{\rm ph}{\rm cm^{2}s~keV}\\\nonumber~~~~~{\rm for}~~E > \frac{(-1.03+2.5)}{(2-1.03)}\times 2.7~{\rm MeV}
\end{eqnarray}
where the normalization constant $C=C_{Band}\sim0.3$ is derived by assuming that the $\nu F_{\nu}$ flux at 100 keV is $\sim 10^{-7}$ erg/cm$^{2}$/s \citep[see the top panel of Fig. 3 in][or the continuous line in our Fig. 1]{Abdoart}, i.e. from Eq. (\ref{Bandlow})
\begin{equation}
(100 \rm{keV})^2 A  \frac{\rm ph}{\rm cm^{2}s~keV} \sim 10^{-7}\frac{\rm{erg}}{\rm cm^2 s}~~~\Rightarrow A\sim(160.2)^{-1}
\end{equation}
and then assuming that the spectrum has a Band shape with $E_{peak}\sim 2.7$ MeV, $\alpha=-1.03$ (as the observed values), and $\beta=-2.5$ \citep[for consistency with the BATSE catalog, as already noticed,  ][]{Kaneko2006}, i.e. from Eq. (\ref{Bandhigh})
\begin{equation}
C_{Band}=\left(\frac{(\alpha-\beta)E_{peak}}{e(2+\alpha)100~\rm{keV}}\right)^{\alpha-\beta} A\sim 0.3.
\end{equation}
Substituting this into Eq. (\ref{gammap}), we thus obtain
\begin{eqnarray}
E_{max}\lesssim 30~{\rm MeV},\\
	\Gamma_a \sim 60 (\delta t_{obs} / 10 {\rm ms})^{-1/7}(E_{max}/30{\rm MeV})^{3/14}.
\end{eqnarray}

 The above equation estimates the Lorentz factor required to keep the shell optically thick to pair production above a few tens of MeVs, as observed during interval a. Detailed modeling of the spectrum expected in the IS scenario for a high compactness shell, is beyond the scope of this paper. As we have seen, optical thickness to pair production affects the observed spectrum at high energies, but when this happens the consequent scattering of photons from the created pairs, and pair annihilation, also need to be taken into account. For instance, when the optical thickness for photon scattering on electrons is high, the spectrum of the observed radiation is modified by the standard assumptions of thin synchrotron and IC emission, and effects related to the so-called electron photosphere need to be considered \citep[see e.g.][]{Meszaros2000}. Re-heating of the electron population caused by synchrotron self-absorption \citep{Ghisellini1988}, is also a process that needs proper evaluation and can modify the spectrum at low energies. Numerical simulations are the most effective way to take into account all these processes dynamically. Within the IS model, the results of detailed numerical modeling by \citet{Peer2004} show that to ensure that the synchrotron emission peaks in the MeV range, as for \object{GRB 081024B}, the required values of the IS model parameters likely imply a high compactness, which causes deviations from the simple predictions of the thin case IS model \citep[e.g.][]{Guetta2003}. For high compactness, \citet{Peer2004} find that the spectra peak at $\sim 1$ MeV, display a steep slope at lower energies (with indices of $0.5\lesssim 2+\alpha\lesssim 1$ in the $\nu F_{\nu}$ spectrum), and a sharp cutoff at $\sim 10$ MeV. This is consistent with the spectrum observed in slice a of GRB081024B, which we therefore attribute to IS emission modified by absorption associated with a high compactness region (in agreement with our analytical estimate). 

\subsection{Interval b: from optically thick to optically thin emission}
During interval a, the source is likely to be optically thick, whereas during interval c, photons with energies of a few GeVs were observed by the Fermi/LAT \citep{Abdoart} thus requiring the source to be optically thin in the GeV range. As discussed in the previous section for interval a, and because the fundamental parameter determining the source compactness is the Lorentz factor of the relativistic shell where the observed radiation is produced, a scenario explaining the observations could be the following. The central engine emits a first shell with Lorentz factor $\Gamma_{a}$, responsible for the emission observed during interval a, with $\Gamma_{a}$ such that the source is optically thick above $\sim 30$ MeV because of the small radius at which the first IS takes place (see previous section). Later on, the central engine emits a series of shells responsible for the other multiple peaks observed during intervals b and c. These shells are characterized by a Lorentz factor in-between $\Gamma_{a}$ and $\Gamma_{c}$, where $\Gamma_{c}>\Gamma_{a}$ is such that the source is thin to GeV photons (as discussed in the next section). 

In the above scenario, spectra observed during interval b and c should follow a progressive transition from an optically thick to an optically thin spectrum in the GeV range. Since both of these intervals contain multiple peaks of the corresponding light curve, we expect that, especially during the transition phase b, the integrated spectrum is a superimposition of spectra emitted by shells with increasing $\Gamma$ factors, progressively more transparent to GeV photons. The best-fit spectrum obtained by Fermi during interval b, does indeed show the contribution from a component peaking around a few MeV (a Band or exponential cutoff component), plus a second component with substantial emission in the GeV range (power-law component). Thus, on general lines, the observed spectral evolution is consistent with our hypothesis. This picture also naturally explains the delayed onset of the GeV tail observed by the LAT. 

\subsection{Interval c: high energy tail from late IS}

\subsubsection{Transparency to GeV photons}
As emphasized before, the observation of GeV photons during interval c requires the Lorentz factor of the late shell generating such emission being sufficiently high for the source to be optically thin at that energies. We thus again use Eq. (\ref{gammap}) assuming that $C=C_{pow}\sim 10^{-4}$, where the $\nu F_{\nu}$ flux at $\sim 2$ GeV is about $10^{-7}$ erg/cm$^{2}$/s \citep[lowest panel of Fig. 3 in][or the dot-dashed line in our Fig. 1]{Abdoart}, and using a photon index of $\beta=-1.6$ (as the observed one), i.e.

\begin{equation}
(2~\rm{GeV})^2 \left(\frac{2~{\rm GeV}}{100~{\rm keV}}\right)^{-1.6} ~C_{pow}~\frac{\rm ph}{\rm cm^{2}s~keV}=10^{-7} \frac{\rm{erg}}{\rm cm^2 s}.
\end{equation}
We also assume that $E_{max}\gtrsim 1{\rm ~GeV}$, and thus

\begin{eqnarray}
E_{max}\gtrsim 2 {\rm GeV},\\
	\Gamma_{c}\sim 70  (\delta t_{obs}/10{\rm ~ms})^{-1/5.2}(E_{max}/2{\rm GeV})^{0.6/5.2},
	\label{reqgamma2}
\end{eqnarray}
so that $\Gamma_c \gtrsim \Gamma_a$. We note that $\Gamma_c$ is not much higher than $\Gamma_a$ because, as can be see from Eq. (\ref{gammap}), in interval c photons of much higher energy are observed (i.e. $2~$ GeV $>> 30$ MeV), although from interval a to c the emitted flux becomes much lower ($C_{pow}<<C_{Band}$).

\subsubsection{GeV emission: synchrotron or SSC?}

In the (optically thin) IS model, the synchrotron peak energy is given by \citep{Guetta2003}

\begin{equation}
\label{piccosincr} E_p=h \nu_m =
1.2\times10^4 \left(\frac{3p-6}{p-1}\right)^{2}
\epsilon_e^{3/2}\epsilon_B^{1/2}L_{52}^{1/2}\Gamma^{-2}(\delta t_{obs}/10~{\rm ms})^{-1}~{\rm
MeV},
\end{equation}
where $L_{52}$ is the source luminosity in units of $10^{52}$ erg \citep{Guetta2003}. We estimate this last parameter by considering the 100 MeV - 10 GeV fluence measured during interval c (and its scatter caused by the measured errors, see Sect. \ref{osservazioni}), and taking into account the observed duration of this interval
\begin{eqnarray}
\nonumber L=4\pi d^2_L \int_0^{\infty}F_\nu d\nu \gtrsim \frac{4\pi d^{2}_L (1+z)~(2-7)\times10^{-7} {\rm erg~cm^{-2}}}{(2.9~{\rm ~s}-0.8~{\rm~s})} \sim\\\sim(3-9)\times10^{48}{~\rm erg~s^{-1}}.~~~~
\label{lumin}
\end{eqnarray}
Using Eq. (\ref{reqgamma2}) into (\ref{piccosincr}), and setting $L_{52}\sim10^{-3}$, we derive

\begin{eqnarray}
	\nonumber E_p\lesssim 0.1\left(\frac{3p-6}{p-1}\right)^{2}
	\epsilon_e^{3/2}\epsilon_B^{1/2}(\delta t_{obs}/10~{\rm ms})^{-0.8/1.3}\times\\\times(E_{max}/2~{\rm GeV})^{-0.3/1.3}~{\rm
	MeV}.
\end{eqnarray}
It is evident that even after setting $\epsilon_e\sim\epsilon_B\sim0.5$, $p\sim5$, and $\delta t_{obs}\sim 1$ ms, the requirement on the Lorentz factor in Eq. (\ref{reqgamma2}) for the source to be optically thin implies that the peak of the synchrotron emission by IS is at energies below $\sim 500$ keV $<< 2$ GeV. We thus conclude that synchrotron emission from late IS cannot explain the high energy tail observed during interval c, since the transparency condition in the GeV range implies values of the synchrotron peak energy much lower than $\sim 1$ GeV, in conflict with the observations.

\begin{figure}
    \begin{center}
        \includegraphics[width=6cm,angle=90]{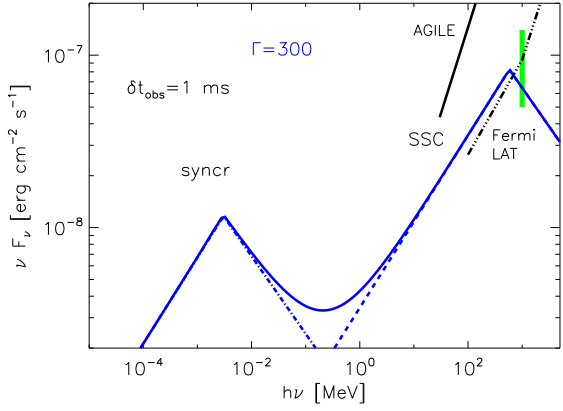}
        \caption{Synchrotron (blue dot-dashed line) and SSC (blue dashed line) emission spectra from delayed IS for a burst with parameters $L_{52}=10^{-3}$, $\delta t_{obs} = 1$ ms, $\epsilon_e= 0.5$, $\epsilon_B = 0.01$, $p = 2.9$, $z=0.1$, $\Gamma=300$. The blue solid line is the sum of the synchrotron and SSC contributions. The black solid and dot-dot-dot-dashed lines are the AGILE and Fermi/LAT sensitivity for an integration time of 10 s, respectively \citep[see ][]{Galli2007}. For this choice of parameters, one has a synchrotron peak around $10$ keV and an SSC peak around $\sim 1$ GeV. The green solid vertical line indicates the flux level at 1 GeV, as reported by \citet{Abdoart} for interval c. See the electronic version of this paper for colors.\label{Fig2}}
    \end{center}
\end{figure}

Another mechanism that may be responsible for the $\sim$ GeV
emission is SSC \citep[see also][]{Zou2008}. 
The peak frequency of the SSC component is given by \citep{Guetta2003}
\begin{eqnarray}
 \nonumber E_p^{SC}=h \nu_m^{SC}= 4.6 \times 10^9
\left(\frac{3p-6}{p-1}\right)^{4}
\epsilon_e^{7/2}\epsilon_B^{1/2}L_{52}^{1/2}\Gamma^{-2}\times\\\times(\delta t_{obs}/10~{\rm ms})^{-1}~{\rm MeV}.\label{piccoSSC}
\end{eqnarray}
Setting $L_{52}=10^{-3}$, $\delta t_{obs} =1$ ms, $\epsilon_e=0.5$, $\epsilon_B = 0.01$, $p = 2.9$, $z=0.1$, into Eq. (\ref{piccoSSC}), we obtain $E_p^{SC}\sim 1$ GeV, for $\Gamma=300$ (see Fig. \ref{Fig2}). For this solution, we note that a variability timescale as short as $1$ ms can indeed be present \citep{Nakar2007}, and was found in at least one short GRB, in which a very bright $< 1$ ms pulse was observed \citep{micropulse}. Moreover, we emphasize that these values of the physical parameters are of course not necessarily unique. However, our aim is to show that a possible solution does indeed exist for a reasonable set of parameters. To derive an order of magnitude estimate of the possible scatter, for each of the parameters we estimated the range into which, leaving the other parameters unchanged, one still obtains $E^{SC}_{p} \gtrsim 100$ MeV, $E_{max} \gtrsim 1$ GeV, and a flux level at 1 GeV compatible with the LAT observations (green vertical line in Fig. 2). In this way, we obtain $p \gtrsim 2.7$, $0.8\times10^{-3} \lesssim L_{52} \lesssim 2\times 10^{-3}$, $0.42\lesssim \epsilon_e \lesssim 0.5$ (where we set the upper limit to ensure that not more than half of the internal energy goes into accelerating the electrons), $2\times10^{-3} \lesssim \epsilon_B \lesssim 0.1$, $0.2 ~{\rm ms} \lesssim \delta t_{obs} \lesssim 2 ~{\rm ms}$, and $140 \lesssim \Gamma \lesssim 430$.

As can be seen in Fig. \ref{Fig2}, in the late IS model the flux level of the high energy tail is within the level measured by the LAT for \object{GRB 081024B} during interval c \citep[see Fig. 3 in][or our Fig. 1]{Abdoart}. Moreover, the predicted $\nu F_{\nu}$ slope below 1 GeV is $1/2$ \citep{Guetta2003}, consistent with the observed value of $2+\left(-1.6^{+0.4}_{-0.1}\right)=0.4^{+0.4}_{-0.1}$ \citep[see Sect. \ref{osservazioni} and][]{Abdoart}. Thus, this model is a viable explanation of the LAT observations of this burst. 

We finally note that more complicated scenarios may be possible, involving a significant contribution at high energies from both the synchrotron and SSC components. For instance, for values of p close to 2, synchrotron emission above the peak results in a flat spectrum, which could be modified by SSC to produce a spectrum similar to the one observed in the case of GRB 081024B. In this case, the GeV emission would not solely be related to SSC, but a significant contribution would come from the synchrotron component as well. 


\section{High energy emission from the ES}

The high energy tail observed in \object{GRB 081024B} (interval c) may also be produced in an extended X-ray tail associated, in this case, with synchrotron afterglow emission by the ES, or alternatively to an afterglow SSC component. We now explore both of these possibilities.

\subsection{Synchrotron-only scenario}
We consider the case in which the high energy tail observed by the Fermi/LAT is the extension to high energies of the synchrotron component generating the afterglow. In this scenario, the afterglow synchrotron emission should match the spectrum observed during interval c \citep[see Sect. \ref{osservazioni} and][]{Abdoart}. To this end, the spectrum should be sufficiently flat to account for an observed photon index of $\beta=-1.6^{+0.4}_{-0.1}$, i.e. $-0.6^{+0.4}_{-0.1}$ in flux. In the case of fast cooling, which is rather natural at such early times, the predicted high energy spectral slope would be $-p/2$ \citep[e.g.][]{Sari1998}, where $p$ is the power-law index of the electron energy distribution behind the shock front. For a typical value of $p\gtrsim 2$, the slope of $-p/2$ would be steeper than observed. On the other hand, in the case of slow cooling, the high energy spectral slope could be $-(p-1)/2\sim -0.6$ for $p=2.2$ \citep[e.g.][]{Sari1998}. 

For $p\sim 2.2$ and slow cooling, the temporal decay of the X-ray light curve would have an index of $-3/4(p-1) \sim -0.9$ \citep{Sari1998}. In an exposure spanning from 70.3 ks to $1.3\times10^{6}$ s after the burst, \textit{Swift}/XRT observed the Fermi-LAT error circle detecting three sources with average count rates below $\sim 2\times10^{-3}$ counts/s \citep{GCNreport}, which we estimate to correspond to an average $0.3-10$~keV flux of $\sim 8\times10^{-14}$~erg~cm$^{-2}$~s$^{-1}$. These sources were excluded as X-ray counterparts of \object{GRB 081024B}, because they did not fade. We can thus use their flux level as an upper limit to the afterglow flux of \object{GRB 081024B} at late times. Using the value of the flux observed at $\sim 2$ GeV during interval c (taking a nominal reference time of 2.5 s after trigger), in a synchrotron-only scenario with $p=2.2$ we expect a 5 keV flux (in the middle of the $0.3-10$~keV XRT band) at 1 day after the burst of about $(5~{\rm keV} /2~{\rm GeV})^{(3-2.2)/2}\times (86400~{\rm s} /2.5~{\rm s})^{-0.9}\times 10^{-7}$~erg~cm$^{-2}$~s$^{-1}\sim 5\times10^{-14}$~erg~cm$^{-2}$~s$^{-1}\lesssim 8\times10^{-14}$~erg~cm$^{-2}$~s$^{-1}$.  This conclusion is valid if the synchrotron cooling frequency is above $\sim 2$ GeV at $\sim 2.5$ s, and above the X-rays one day after the burst. \citet{He2009} interpreted the data of \object{GRB~081024B} in the slow cooling case as well. We thus conclude that the Fermi/LAT observations may be explained as synchrotron emission from an early FS afterglow, if the slow cooling regime occurs as soon as $\sim 2$ s after the trigger. We note that a synchrotron ES scenario has also been proposed to explain the high energy tail observed in GRB 090510 \citep[e.g.][]{DePasquale2010,Ghirlanda2009,Corsi2009}.


\subsection{Synchrotron plus SSC scenario}

We can alternatively link the high energy tail associated with \object{GRB 081024B}, to an ES SSC component entering into the observed band (while the ES synchrotron emission is shifted to lower energies and lower fluxes). In this case, if the SSC peak were around 1 GeV, the predicted spectral index below 1 GeV would be in the range of $[-1/2,1/3]$ \citep[e.g.][]{SariEsin2001}, which should be compared with the observed value of $1+\beta=-0.6^{+0.4}_{-0.1}$ \citep{Abdoart}. We now analyze this scenario in more detail.

\subsubsection{Synchrotron component}

Following the prescriptions by \citet{SariEsin2001}, we can
express the characteristic break frequencies and the peak flux
of the synchrotron component as

\begin{equation}
    \nu_m =5\times10^{12}~{\rm Hz}~(1+z)^{1/2}\frac{f(p)}{f(2.2)}\left(\frac{\epsilon_{B}}{0.01}\right)^{1/2}\left(\frac{\epsilon_{e}}{0.5}\right)^2 E_{52}^{1/2} t_{day}^{-3/2}
    \label{prima}
    \end{equation}

where $f(p)=((p-2)/(p-1))^{2}$,

\begin{equation}
    \nu_c =\frac{2.7\times10^{15}}{(1+z)^{1/2}}~{\rm Hz}~ \left(\frac{\epsilon_{B}}{0.01}\right)^{-3/2} E_{52}^{-1/2}n_1^{-1} t_{day}^{-1/2} (1+Y)^{-2},
    \label{primac}
\end{equation}

\begin{equation}
f_{m}=2.6~{\rm mJy}~(1+z)
\left(\frac{\epsilon_{B}}{0.01}\right)^{1/2} E_{52}~n_1^{1/2}
d_{L,28}^{-2}.~~
\end{equation} 
As in the previous sections, $Y=\frac{L_{IC}}{L_{syn}}$, and
\textit{in the fast cooling regime} $Y\sim
\sqrt{\frac{\epsilon_e}{\epsilon_B}}$ \cite[e.g.
][]{SariEsin2001}. \textit{In this regime} the energy spectrum
$\nu F_{\nu}$ peaks at $\nu_m$, thus $Y \sim (\nu^{IC}_m
f^{IC}(\nu^{IC}_m))/(\nu_m f(\nu_m))$, where

\begin{eqnarray}
    \nonumber L^{syn}=\nu_m (\nu_m/\nu_c)^{-1/2}f_{m}= 4.3\times10^{-13}~{\rm \frac{erg}{~cm^{2}~s^{1}}}\times\\\times\left(\frac{f(p)}{f(2.2)}\right)^{1/2}(1+z) \left(\frac{\epsilon_{e}}{0.5}\right)^{1/2}\left(\frac{\epsilon_{B}}{0.01}\right)^{1/2} E_{52} t_{day}^{-1} d^{-2}_{L,28}~~~\label{seconda}
\end{eqnarray}
and we have used Eqs. (\ref{prima})-(\ref{primac}), and $(1+Y)^{-2}\sim
Y^{-2}=\epsilon_B/\epsilon_e$. If the peak of the synchrotron
component in the $\nu F_{\nu}$ space is below $1$~keV, i.e. if
$\nu_m<1$~keV, we can substitute $L^{syn}$ on the left hand side
of the above equation with the expression

\begin{eqnarray}
L^{syn}=F^{syn}_{1 keV}\left({\rm
\frac{erg}{cm^{2}~s~Hz}}\right)(2.41\times10^{17}~{\rm
Hz})^{p/2}(\nu_m({\rm Hz}))^{-p/2+1}.~~~ \label{terza}
\end{eqnarray}
In this way, from Eqs. (\ref{prima}) and
(\ref{seconda})-(\ref{terza}), we derive the expressions
for $\epsilon_e$ and $\epsilon_B$ of

\begin{eqnarray}
\nonumber
\epsilon_{B}=\frac{0.2}{(1+z)^{7/3}}~\left(\frac{f(p)}{f(2.2)}\right)^{-2/3}\left(\frac{\nu_{m}}{1\rm
keV}{}\right)^{-4p/3+2}\times\\\times\left(\frac{F^{syn}_{1 \rm
keV}}{10~{\rm mJy}}\right)^{8/3} \left(\frac{t_{\rm
s}^{5}d^{16}_{L,28}}{E_{52}^{7}}\right)^{1/3},\label{quarta}
\end{eqnarray}

\begin{eqnarray}
\nonumber \epsilon_{e}=0.01~\left(\frac{f(p)}{f(2.2)}\right)^{-1/3}\left(\frac{\nu_{m}}{1\rm
keV}{}\right)^{p/3}\left(\frac{F^{syn}_{1 \rm keV}}{10~{\rm
mJy}}\right)^{-2/3}\times\\\times\left(\frac{(1+z)E_{52}t_{\rm
s}}{d^{4}_{L,28}}\right)^{1/3}.\label{quinta}
\end{eqnarray}
The above equations allow us to eliminate from the problem the two unknown micro-physical
parameters by expressing them as a function of the synchrotron peak frequency $\nu_m$ and the
observed $1$ keV flux. We estimate the typical X-ray luminosity of a short GRB by considering the
$0.3-10$~keV fluxes at $100$~s, $F_{0.3-10 keV,~100~s}$, reported in Table 2 of
\citet{Nakar2007}, which are in between $6\times10^{-13}$~erg~cm$^{-2}$~s$^{-1}$ and
$1.2\times10^{-8}$~erg~cm$^{-2}$~s$^{-1}$, with a mean value of $<F_{0.3-10~{\rm
keV},~100~{\rm s}}>\simeq 2\times10^{-9}$~erg~cm$^{-2}$~s$^{-1}$. For $p=2.05$ (so as to
favor the emission at high energies by having a flat spectrum), we can thus estimate $F_{1~{\rm
keV},~2.5~{\rm s}}$ by using a spectral slope of $-p/2\sim -1$ in the $0.3-10$ keV
range (i.e. assuming that $\nu_m\lesssim0.3$ keV), and a temporal decay index of
$-3/4(p-1)-1/4\sim -1$. Doing so, we find that $F_{1~keV,~2.5~s}\sim 10$~mJy is a
reasonable estimate. To constrain $E_{52}$, as done in Eq. (\ref{lumin}), we estimate
$E_{52}\gtrsim E_{\gamma, 52}=2\times10^{-3}$ at $z=0.1$.


\subsubsection{IC component}

\begin{figure}
    \begin{center}
        \includegraphics[width=6.5cm,angle=90]{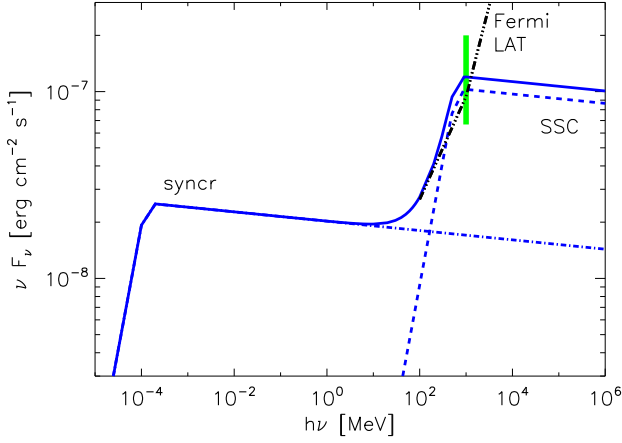}
        \caption{Synchrotron (dot-dashed line) and SSC (dashed line) spectra in the
        ES scenario,
         for a burst with parameters $p=2.05$, $E_{52}=0.35$, $z=0.1$,
         $n_1=5$, $\nu_m=0.15$ keV, $F^{syn}_{\rm 1 keV}=10$ mJy, and $t_{\rm s}=2.5$.
         The solid line is the sum of the synchrotron and SSC contributions.
         For this choice of parameters, one has an SSC peak around $\sim 1$ GeV.
         The dot-dot-dot-dashed line represents the Fermi/LAT sensitivity for an integration time of 10 s \citep[see ][]{Galli2007}. The green solid vertical line marks the observed flux level at 1 GeV, as reported by \citet{Abdoart} for interval c. See the electronic version of this paper for colors.
        \label{ES}}
    \end{center}
\end{figure}

In the fast cooling regime, the IC energy emission peaks at

\begin{eqnarray}
\nonumber   \nu^{IC}_m =2\gamma^2_m \nu_m=3.7~{\rm
GeV}\left(\frac{f(p)}{f(2.2)}\right)^{1/3}
\left(\frac{\nu_{m}}{1\rm
keV}\right)^{2p/3+1}\times\\\times\left(\frac{F^{syn}_{1 \rm keV}}{10~{\rm
mJy}}\right)^{-4/3} E_{52}^{11/12} n_1^{-1/4}t_{\rm
s}^{-1/12} (1+z)^{17/12} d^{-8/3}_{L,28}~~ \label{settima}
\end{eqnarray}
where we have used \citep{SariEsin2001}:

\begin{equation}
\gamma_m=930 \left(\frac{f(p)}{f(2.2)}\right)^{1/2}
\left(\frac{\epsilon_e}{0.5}\right)\left(\frac{E_{52}}{n_1}\right)^{1/8}
\left(\frac{t_{day}}{1+z}\right)^{-3/8}
\end{equation}
with Eqs. (\ref{quarta})-(\ref{quinta}). Setting
$p=2.05$, $E_{52}=0.35$, $z=0.1$, $n_1=5$, $\nu_m=0.15$ keV,
$F^{syn}_{\rm 1 keV}=10$ mJy, and $t=2.5$ s in the above
equation, we derive $\nu^{IC}_{m}\sim 1$ GeV (see Fig. \ref{ES}). We note that $E_{52}=0.35$, compared to the value of $E_{\gamma, 52}=2\times10^{-3}$ estimated from the prompt and high energy tail fluence, implies that the conversion efficiency into $\gamma$-rays is $\sim 1\%$, which is at the lower end of the typical range $0.01-1$ found for long GRBs and probably the same for short GRBs \cite[see e.g. ][]{Nakar2007,Zhang2007}. The
IC flux at the peak $\nu^{IC}_m$ is given by
\begin{eqnarray}
    \nonumber \nu^{IC}_{m}f^{IC}({\nu^{IC}_m})=Y L^{syn}=5.3\times10^{-9}{\frac{\rm erg}{\rm cm^{2}s}}\left(\frac{f(p)}{f(2.2)}\right)^{1/6}\times\\\times\left(\frac{\nu_{m}}{{1~\rm keV}}\right)^{p/3}\left(\frac{F^{syn}_{\rm 1 keV}}{10~{\rm mJy}}\right)^{-2/3}(1+z)^{4/3} E^{4/3}_{52} t^{-2/3}_{\rm s} d^{-10/3}_{L,28},~~~~
\end{eqnarray}
where we have used Eqs. (\ref{seconda}),
(\ref{quarta})-(\ref{quinta}). For the same set of parameters, we
have $\nu^{IC}_{m}f^{IC}({\nu^{IC}_m})\sim 10^{-7}$~erg~cm$^{-2}$~s$^{-1}$ (see
Fig. \ref{ES}), which is comparable with the LAT sensitivity for
$10$ s integration time. We note that for a given value of $\nu_m$, $F^{syn}_{\rm 1 keV}$, and $z$,
the above equation ensures that $E_{52}$ is sufficiently high to have the GeV tail detected
by the Fermi/LAT. At the same time, it is evident from Eq. (\ref{settima}) that a higher value of $E_{52}$ tends to shift the peak energy to higher values, so that to keep it around $\sim 1$ GeV, $n$ cannot be too low. Our value of $n=5$~cm$^{-3}$
is in the range that has been found to possibly characterize other short bursts \citep[see e.g. ][]{Panaitescu2005}, and roughly at the higher edge of the $0.01-1$~cm$^{-3}$ range expected for the ISM.

\subsubsection{Consistency checks: micro-physics, deceleration/cooling time, and Klein-Nishina limit}

To determine whether the ES scenario proposed in this section is a
self-consistent explanation of the high energy tail observed in
\object{GRB 081024B}, we need to perform a series of checks to
verify that the hypotheses under which we operate are
consistent with our choice of parameters. First of all,
the inferred values of $\epsilon_e$ and $\epsilon_B$ should both be
less than unity, and we should have $\epsilon_e>>\epsilon_B$. With
our choice of parameters, we find that $\epsilon_B=5.2\times10^{-3}$ and
$\epsilon_e=8.8\times10^{-2}$, which are consistent with these conditions. 

To have an ES, we need the deceleration phase to begin before or around the time at which the high energy tail
is observed, i.e., \citep{Sari1999}
\begin{equation}
    t_{dec}\sim 3.2~{\rm s}~\left(\frac{E_{52}}{n_1}\right)^{1/3}\left(\frac{\Gamma_0}{350}\right)^{-8/3} (1+z)\lesssim 2.5 {\rm s},
    \label{tdec}
\end{equation}
which for $z=0.1$, $E_{52}=0.35$, and $n_1=5$ implies that
$\Gamma_0\gtrsim 285$, which is a reasonable lower-limit
to the initial fireball Lorentz factor.
We note that, although (as shown here) the minimum mathematical condition for having a deceleration time as early as few seconds does indeed hold, Swift seems to detect \textit{long} GRBs with complex early-time light curves, and clear evidence of a self-similar motion on a timescale only of 1000 s or longer. Thus, on the very short timescale considered here, more complex hydrodynamics (such  as e.g. a reverse shock or energy injection) may occur, producing a complex light curve. On the other hand, we emphasize that a complex light curve behavior at early times is not as evident for \textit{short} GRBs, as it is for \textit{long} ones observed by Swift. Moreover, Fermi observations of short GRBs associated with high energy tails, as for GRB 090510 \citep{Abdo2009,Giuliani2009} indeed appear to detect a smooth high energy light curve, with evidence for the fireball entering in the self-similar phase as early as a few seconds after the burst \citep{Ghirlanda2009}.  We thus consider the scenario described in this section to be a simple, but realistic and viable description of the physical processes relevant to the short \object{GRB~081024B}.

We have also considered the hypothesis of 
the fast cooling regime, which we need to verify is indeed
the case, i.e. $\nu_c(2.5 {\rm s}) < \nu_m(2.5 {\rm s})$. Using Eq. (\ref{primac}) we find that 
$\nu_c(2.5 {\rm s})\sim 0.1$ keV, so the fast cooling hypothesis is applicable, and the
fast-to-slow cooling transition occurs at about $3.6$ s after the burst.




Finally, we need to check
that the Klein-Nishina effect does not suppress the IC component.
In the fast cooling regime, most of the synchrotron energy is
emitted around $\nu_m$ and most of the SSC energy is emitted by
electrons with $\gamma_e\sim\gamma_m$ that up-scatter photons with
$\nu\sim\nu_m$. Therefore, the Klein-Nishina limit can be
neglected only if \citep[\textbf{see e.g.}][]{Rib1986}

\begin{equation}
    \nu_m\lesssim \nu_{KN}(\gamma_m)=\frac{m_e c^2 \Gamma}{\gamma_m}
\end{equation}
Since $\gamma_m=(m_p/m_e)(f(p))^{1/2}\epsilon_e \Gamma$ \citep[][]{Sari1998},
this condition implies that

\begin{equation}
\left(\frac{\nu_m}{1~\rm keV}\right)\lesssim 3.3
\left(\frac{\epsilon_{e}}{0.5}\right)^{-1} \left(\frac{f(p)}{f(2.2)}\right)^{-1/2}
\end{equation}
In our case we have $\nu_m=0.15$ keV at $t=2.5$ s, while the
right-hand side of the above equation computed for
$\epsilon_e=8.8\times10^{-2}$ and $p=2.05$ is equal to $\sim 66$
keV. Thus, the above condition is also verified.

\section{Discussion and conclusion}

We have investigated the origin of both the prompt emission and high-energy
tail associated with \object{GRB 081024B}, by exploring four main scenarios:

\begin{enumerate}
		\item synchrotron IS emission (first main peak, interval a);
    \item (synchrotron or) SSC component associated with a delayed X-ray emission produced by late IS (high-energy tail);
    \item synchrotron component from the ES generating the afterglow emission (high energy tail);
    \item a SSC component from the ES generating the afterglow emission (high-energy tail).
\end{enumerate}

To derive the model parameters, we have considered the
observational constraints provided by the analysis by \citet{Abdoart}.
By comparing with previous studies, we have confirmed the results by \citet{Zou2008}, which we have expanded in the following way. While in \citet{Zou2008}, the late IS SSC scenario was restricted to noting that the SSC peak frequency can be $\sim 100$ MeV for reasonable parameter values, here we have shown that solutions can be found that also satisfy two additional constraints: (a) the source is optically thin around 1 GeV; (b) the flux level at 1 GeV is compatible with that observed by Fermi LAT. The discussion about the SSC from the ES scenario in \citet{Zou2008} was also restricted to noting that the SSC peak frequency may be in the GeV range for reasonable parameter values. Here we have shown that a reasonable set of parameters can be found that also implies a flux level at 1 GeV compatible with the one observed by Fermi LAT. Moreover, we have considered two additional scenarios (1. and 3.).

We have shown that scenarios 2. (SSC) and 4. are viable explanations of
the observed tail for a burst located at $z\sim 0.1$. To reproduce the high energy tail in a delayed IS scenario, the lately emitted shells should have a time variability of about $1$ ms and a Lorentz factor
of about $\Gamma=300$. In the ES shock scenario, the high energy
tail can be explained by assuming a flat spectrum, i.e. $p=2.05$,
and that the short GRB is powered by a fireball with an
isotropic energy of about $10^{51}$ erg, expanding in an ISM with density $n=5$ cm$^{-3}$.
These values of the parameters are order-of-magnitude estimates due to the uncertainties in the early-time afterglow flux, which was not observed for this burst. In particular, the fast cooling condition ($t_{cool}\gtrsim 2.5$ s), which is reasonable to expect at the early times we consider here, depends linearly on the chosen value of $n$ and almost linearly on the early-time afterglow flux value. Equating Eq. (\ref{prima}) to Eq. (\ref{primac}), indeed one finds that $t_{cool}\propto n F_{1 \rm keV}^{2/3}$, so that a value of $n$ in the lower end of the range of values expected for short GRBs would require a higher value of $F_{1 \rm keV}$ to ensure that $t_{cool}\gtrsim 2.5$ s \cite[see e.g.][for an alternative interpretation in the case of slow cooling]{He2009}. These  estimates, however, are the most robust that can be derived from the publicly available data. They are also sufficient to show that a solution does indeed exist for a reasonable set of parameters, which is the aim of this work.
We emphasize that scenarios 2, 3, and 4, which are related to the emission from a lately emitted shell (2) or from the ES deceleration phase (3 and 4), all offer a natural explanation of the observed temporal delay between the high energy tail and the main burst. Moreover, scenario 2 (emission from a lately emitted shell) may be consistent with the steeply declining emission from an extended X-ray tail that has been observed in association with some short GRBs before 100 s after the trigger time \citep[see Fig. 7 in][]{Nakar2007}, rather than the ``normal'' decay typical of the afterglow emission related to a decelerating ES.

Finally, we have also underlined that other explanations may exist, e.g. where the initial lack of GeV photons is due to a fireball dominated by emission from a photosphere, rather than from an absorbed synchrotron spectrum \citep[see e.g. ][and references therein]{Ioka2010,Mizuta2010,Peer2010,Toma2010}.

\begin{acknowledgements}
We thank the anonimous Referee of this paper for providing very useful comments on our work.
We thank Alessandra Galli for valuable discussions. A. Corsi is grateful to the Italian
L'Oreal-UNESCO program ``For Women in Science'' for support. The authors acknowledge the
support of ASI-INAF contract I/088/06/0 and of EGO - European Gravitational Wave Observatory.
\end{acknowledgements}

\bibliographystyle{aa}

\bibliography{12461}

\begin{thebibliography}{72}
\expandafter\ifx\csname natexlab\endcsname\relax\def\natexlab#1{#1}\fi

\bibitem[{{Abdo} {et~al.}(2009{\natexlab{a}})}]{science}
{Abdo}, A.~A. {et~al.} 2009{\natexlab{a}}, Science, 323, 1688

\bibitem[{{Abdo} {et~al.}(2009{\natexlab{b}})}]{Abdo2009}
{Abdo}, A.~A. {et~al.} 2009{\natexlab{b}}, Nature, 462, 331

\bibitem[{{Abdo} {et~al.}(2010)}]{Abdoart}
{Abdo}, A.~A. {et~al.} 2010, ApJ, 712, 558

\bibitem[{{Asano} \& {Inoue}(2007)}]{Asano2007}
{Asano}, K. \& {Inoue}, S. 2007, ApJ, 671, 645

\bibitem[{{Band} {et~al.}(1993)}]{Band1993}
{Band}, D. {et~al.} 1993, ApJ, 413, 281

\bibitem[{{Baring} \& {Braby}(2004)}]{Baring2004}
{Baring}, M.~G. \& {Braby}, M.~L. 2004, ApJ, 613, 460

\bibitem[{{Beloborodov}(2000)}]{Beloborodov2000}
{Beloborodov}, A.~M. 2000, \apjl, 539, L25

\bibitem[{{Chandra} {et~al.}(2008)}]{Chandra2008}
{Chandra}, P. {et~al.} 2008, ApJ, 683, 924

\bibitem[{{Corsi} {et~al.}(2010){Corsi}, {Guetta}, \& {Piro}}]{Corsi2009}
{Corsi}, A., {Guetta}, D., \& {Piro}, L. 2010, ApJ, 720, 1008

\bibitem[{{Corsi} \& {Piro}(2006)}]{Corsi2006}
{Corsi}, A. \& {Piro}, L. 2006, A\&A, 458, 741

\bibitem[{{Corsi} {et~al.}(2005)}]{Corsi2005}
{Corsi}, A. {et~al.} 2005, A\&A, 438, 829

\bibitem[{{Dai} \& {Lu}(2002)}]{DaiLu2002}
{Dai}, Z.~G. \& {Lu}, T. 2002, ApJ, 580, 1013

\bibitem[{{Daigne} \& {Mochkovitch}(1998)}]{Daigne1998}
{Daigne}, F. \& {Mochkovitch}, R. 1998, \mnras, 296, 275

\bibitem[{{De Pasquale} {et~al.}(2010)}]{DePasquale2010}
{De Pasquale}, M. {et~al.} 2010, ApJL, 709, L146

\bibitem[{{Fan}(2009)}]{Fan09}
{Fan}, Y. 2009, MNRAS, 397, 1539

\bibitem[{{Fan} \& {Piran}(2008)}]{Fan2008b}
{Fan}, Y.-Z. \& {Piran}, T. 2008, Front. Phys. China, 3, 306

\bibitem[{{Fan} {et~al.}(2008)}]{Fan2008}
{Fan}, Y.-Z. {et~al.} 2008, MNRAS, 384, 1483

\bibitem[{{Galli} \& {Piro}(2007)}]{Galli2007}
{Galli}, A. \& {Piro}, L. 2007, A\&A, 475, 421

\bibitem[{{Galli} \& {Piro}(2008)}]{Galli2008}
{Galli}, A. \& {Piro}, L. 2008, A\&A, 489, 1073

\bibitem[{{Galli} \& {Guetta}(2008)}]{GalliGuetta2008}
{Galli}, G. \& {Guetta}, D. 2008, A\&A, 480, 5

\bibitem[{{Ghirlanda} {et~al.}(2010){Ghirlanda}, {Ghisellini}, \&
  {Nava}}]{Ghirlanda2009}
{Ghirlanda}, G., {Ghisellini}, G., \& {Nava}, L. 2010, A\&A, 510, L7

\bibitem[{{Ghisellini} {et~al.}(2000){Ghisellini}, {Celotti}, \&
  {Lazzati}}]{Ghisellini2000}
{Ghisellini}, G., {Celotti}, A., \& {Lazzati}, D. 2000, MNRAS, 313, 1

\bibitem[{{Ghisellini} {et~al.}(1988){Ghisellini}, {Guilbert}, \&
  {Svensson}}]{Ghisellini1988}
{Ghisellini}, G., {Guilbert}, P.~W., \& {Svensson}, R. 1988, ApJL, 334, L5

\bibitem[{{Giuliani} {et~al.}(2010)}]{Giuliani2009}
{Giuliani}, A. {et~al.} 2010, ApJL, 708, L84

\bibitem[{{Gou} \& {M\'esz\'aros}(2007)}]{Gou2007}
{Gou}, L.-J. \& {M\'esz\'aros}, P. 2007, ApJ, 668, 392

\bibitem[{{Guetta} \& {Granot}(2003)}]{Guetta2003}
{Guetta}, D. \& {Granot}, J. 2003, ApJ, 585, 885

\bibitem[{{Guidorzi} {et~al.}(2008{\natexlab{a}})}]{GCN8410}
{Guidorzi}, C. {et~al.} 2008{\natexlab{a}}, GCN, 8410

\bibitem[{{Guidorzi} {et~al.}(2008{\natexlab{b}})}]{GCN8416}
{Guidorzi}, C. {et~al.} 2008{\natexlab{b}}, GCN, 8416

\bibitem[{{Guidorzi} {et~al.}(2008{\natexlab{c}})}]{GCN8454}
{Guidorzi}, C. {et~al.} 2008{\natexlab{c}}, GCN, 8454

\bibitem[{{Guidorzi} {et~al.}(2008{\natexlab{d}})}]{GCNreport}
{Guidorzi}, C. {et~al.} 2008{\natexlab{d}}, GCN report, 178

\bibitem[{{Hanabata} {et~al.}(2008)}]{GCN8444}
{Hanabata}, Y. {et~al.} 2008, GCN, 8444

\bibitem[{{Harrison} {et~al.}(2001)}]{Harrison2001}
{Harrison}, F.~A. {et~al.} 2001, ApJ, 559, 123

\bibitem[{{He} \& {Wang}(2009)}]{He2009}
{He}, H.-N. \& {Wang}, X.-Y. 2009, ArXiv e-prints, 0908.2580

\bibitem[{{Ioka}(2010)}]{Ioka2010}
{Ioka}, K. 2010, Progress of Theoretical Physics, 124, 667

\bibitem[{{Kaneko} {et~al.}(2006){Kaneko}, {Preece}, {Briggs}, {Paciesas},
  {Meegan}, \& {Band}}]{Kaneko2006}
{Kaneko}, Y., {Preece}, R.~D., {Briggs}, M.~S., {et~al.} 2006, ApJS, 166, 298

\bibitem[{{Li}(2010)}]{Li2008}
{Li}, Z. 2010, ApJ, 709, 525

\bibitem[{{Lithwick} \& {Sari}(2001)}]{Lithwick2003}
{Lithwick}, Y. \& {Sari}, R. 2001, ApJ, 555, 540

\bibitem[{{M\'esz\'aros} \& {Rees}(1992)}]{Meszaros1993}
{M\'esz\'aros}, P. \& {Rees}, M. 1992, MNRAS, 258, 41

\bibitem[{{M{\'e}sz{\'a}ros} \& {Rees}(2000)}]{Meszaros2000}
{M{\'e}sz{\'a}ros}, P. \& {Rees}, M.~J. 2000, \apj, 530, 292

\bibitem[{{Mizuta} {et~al.}(2010){Mizuta}, {Nagataki}, \& {Aoi}}]{Mizuta2010}
{Mizuta}, A., {Nagataki}, S., \& {Aoi}, J. 2010, ArXiv e-prints, 1006.2440

\bibitem[{{Nakar}(2007)}]{Nakar2007}
{Nakar}, E. 2007, PhR, 442, 166

\bibitem[{{Omodei}(2008)}]{GCN8407}
{Omodei}, N. 2008, GCN, 8407

\bibitem[{{Panaitescu}(2006)}]{Panaitescu2005}
{Panaitescu}, A. 2006, MNRAS, 367, L42

\bibitem[{{Panaitescu} \& {M\'esz\'aros}(2000)}]{PanaitescuMes2000}
{Panaitescu}, A. \& {M\'esz\'aros}, P. 2000, ApJ, 544, 17

\bibitem[{{Panaitescu} {et~al.}(1999){Panaitescu}, {Spada}, \&
  {M{\'e}sz{\'a}ros}}]{Panaitescu1999}
{Panaitescu}, A., {Spada}, M., \& {M{\'e}sz{\'a}ros}, P. 1999, \apjl, 522, L105

\bibitem[{{Papathanassiou} \& {M\'esz\'aros}(1996)}]{Pa1996}
{Papathanassiou}, H. \& {M\'esz\'aros}, P. 1996, ApJ, 471, L91

\bibitem[{{Pe'er} \& {Ryde}(2010)}]{Peer2010}
{Pe'er}, A. \& {Ryde}, F. 2010, ArXiv e-prints, 1003.2582

\bibitem[{{Pe'er} \& {Waxman}(2004)}]{Peer2004}
{Pe'er}, A. \& {Waxman}, E. 2004, ApJ, 613, 448

\bibitem[{{Pe'er} \& {Waxman}(2005)}]{Asaf2005}
{Pe'er}, A. \& {Waxman}, E. 2005, ApJ, 633, 1018

\bibitem[{{Pilla} \& {Loeb}(1998)}]{Pilla1998}
{Pilla}, R.~P. \& {Loeb}, A. 1998, ApJ, 494, L167

\bibitem[{{Piran}(1999)}]{Piran1999}
{Piran}, T. 1999, Phys. Rep., 314, 575

\bibitem[{{Ramirez-Ruiz} \& {Fenimore}(2000)}]{Ramirez2000}
{Ramirez-Ruiz}, E. \& {Fenimore}, E.~E. 2000, \apj, 539, 712

\bibitem[{{Rees} \& {Meszaros}(1994)}]{Meszaros1994}
{Rees}, M.~J. \& {Meszaros}, P. 1994, ApJL, 430, L93

\bibitem[{{Rybicki} \& {Lightman}(1986)}]{Rib1986}
{Rybicki}, G.~B. \& {Lightman}, A.~P. 1986, {Radiative Processes in
  Astrophysics}, ed. {Rybicki, G.~B.~\& Lightman, A.~P., Wiley-VCH}

\bibitem[{{Sari} \& {Esin}(2001)}]{SariEsin2001}
{Sari}, R. \& {Esin}, A.~A. 2001, ApJ, 548, 787

\bibitem[{{Sari} \& {Piran}(1997)}]{Sari1997}
{Sari}, R. \& {Piran}, T. 1997, MNRAS, 287, 110

\bibitem[{{Sari} \& {Piran}(1999)}]{Sari1999}
{Sari}, R. \& {Piran}, T. 1999, ApJ, 520, 641

\bibitem[{{Sari} {et~al.}(1996){Sari}, {Piran}, \& {Narayan}}]{sari96}
{Sari}, R., {Piran}, T., \& {Narayan}, R. 1996, ApJ, 473, 204

\bibitem[{{Sari} {et~al.}(1998){Sari}, {Piran}, \& {Narayan}}]{Sari1998}
{Sari}, R., {Piran}, T., \& {Narayan}, R. 1998, ApJ, 497, L17

\bibitem[{{Scargle} {et~al.}(1998){Scargle}, {Norris}, \&
  {Bonnell}}]{micropulse}
{Scargle}, J.~D., {Norris}, J.~P., \& {Bonnell}, J.~T. 1998, AIP Conference
  Proceedings, 428, 181

\bibitem[{{Spada} {et~al.}(2000){Spada}, {Panaitescu}, \&
  {M{\'e}sz{\'a}ros}}]{Spada2000}
{Spada}, M., {Panaitescu}, A., \& {M{\'e}sz{\'a}ros}, P. 2000, \apj, 537, 824

\bibitem[{{Svensson}(1987)}]{Svensson1987}
{Svensson}, R. 1987, \mnras, 227, 403

\bibitem[{{Toma} {et~al.}(2010){Toma}, {Wu}, \& {Meszaros}}]{Toma2010}
{Toma}, K., {Wu}, X., \& {Meszaros}, P. 2010, ArXiv e-prints, 1002.2634

\bibitem[{{Wang} {et~al.}(2006){Wang}, {Li}, \& {M\'esz\'aros}}]{Wang2006}
{Wang}, X.-Y., {Li}, Z., \& {M\'esz\'aros}, P. 2006, ApJ, 641, 89

\bibitem[{{Wang} {et~al.}(2009){Wang}, {Dai}, \& {M\'esz\'aros}}]{Wang2009}
{Wang}, X.-Y.~Z., {Dai}, Z.~G., \& {M\'esz\'aros}, P. 2009, ApJ, 698, L98

\bibitem[{{Wei} \& {Lu}(1998)}]{Wei1998}
{Wei}, D.~M. \& {Lu}, T. 1998, ApJ, 505, 252

\bibitem[{{Wei} \& {Lu}(2000)}]{Wei2000}
{Wei}, D.~M. \& {Lu}, T. 2000, A\&A, 360, 13

\bibitem[{{You} {et~al.}(2007){You}, {Liu}, \& {Dai}}]{You2007}
{You}, Y.~W., {Liu}, X.~W., \& {Dai}, Z.~G. 2007, ApJ, 671, 637

\bibitem[{{Yu} \& {Dai}(2009)}]{Yu2009}
{Yu}, Y.~W. \& {Dai}, Z.~G. 2009, ApJ, 692, 133

\bibitem[{{Zhang} \& {M\'esz\'aros}(2001)}]{ZhangMes2001}
{Zhang}, B. \& {M\'esz\'aros}, P. 2001, ApJ, 559, 110

\bibitem[{{Zhang} {et~al.}(2007)}]{Zhang2007}
{Zhang}, B. {et~al.} 2007, ApJ, 655, 989

\bibitem[{{Zou} {et~al.}(2009){Zou}, {Fan}, \& {Piran}}]{Zou2008}
{Zou}, Y.-C., {Fan}, Y.-Z., \& {Piran}, T. 2009, MNRAS, 396, 1163

\end{thebibliography}

\end{document}